\documentclass[12pt]{iopart}
\usepackage{graphicx}
\usepackage{cite}
\begin{document}

\title[Impact of backscattered light in a squeezed interferometric gravitational-wave detector]{Impact of backscattered light in a squeezing-enhanced interferometric gravitational-wave detector}

\author{S S Y Chua,$^{1}$ S Dwyer,$^{2}$ L Barsotti,$^{2}$ D Sigg,$^{3}$ R M S Schofield,$^{4}$ V V Frolov,$^{5}$ K Kawabe,$^{3}$ M Evans,$^{2}$ G D Meadors,$^{6}$ M Factourovich,$^{7}$ R Gustafson,$^{6}$ N Smith-Lefebvre,$^{2}$ C Vorvick,$^{3}$ M Landry,$^{3}$ A Khalaidovski,$^{8}$ M S Stefszky,$^{1}$ C M Mow-Lowry,$^{1}$ B C Buchler,$^{1}$ D A Shaddock,$^1$ P K Lam,$^1$ R Schnabel,$^{8}$ N Mavalvala$^{2}$ and D E McClelland.$^{1}$}

\address{
$^1$Department of Quantum Science, The Australian National University, ACT 0200, Australia\\
$^2$LIGO Laboratory, Massachusetts Institute of Technology, Cambridge, MA 02139, USA\\
$^3$LIGO Hanford Observatory, PO Box 159, Richland, WA 99352, USA\\
$^4$University of Oregon, Eugene, OR 97403, USA\\
$^5$LIGO Livingston Observatory, PO Box 940, Livingston, LA 70754, USA\\
$^6$University of Michigan, Ann Arbor, MI 48109, USA\\
$^7$Columbia University, New York, NY 10027, USA\\
$^8$Institut f\"{u}r Gravitationsphysik of Leibniz Universit\"{a}t Hannover and Max-Planck-Institut f\"{u}r
Gravitationsphysik (Albert-Einstein-Institut), Callinstr. 38, 30167 Hannover, Germany}
\ead{sheon.chua@anu.edu.au}

\begin{abstract}
Squeezed states of light have been recently used to improve the sensitivity of laser interferometric gravitational-wave detectors beyond the quantum limit. To completely establish quantum engineering as a realistic option for the next generation of detectors, it is crucial to study and quantify the noise coupling mechanisms which injection of squeezed states could potentially introduce. We present a direct measurement of the impact of backscattered light from a squeezed-light source deployed on one of the 4 km long detectors of the Laser Interferometric Gravitational Wave Observatory (LIGO). We also show how our measurements inform the design of squeezed light sources compatible with the even more sensitive advanced detectors currently under construction, such as Advanced LIGO.
\end{abstract}

\submitto{\CQG}
\maketitle

Laser-interferometric gravitational wave detectors, such as those of the Laser Interferometer Gravitational Wave Observatory (LIGO), are the most sensitive position meters yet made, able to measure length variations of order $10^{-19}$ m over a multi-kilometer baseline. The Advanced LIGO detectors currently under construction aim to achieve even greater sensitivities, of the order of $10^{-20}~\rm{m/ \sqrt{Hz}}$ at 200 Hz \cite{Harry}.

The LIGO interferometers are limited by quantum noise down to 150 Hz, and the Advanced LIGO detectors are expected to be limited by quantum noise across their entire measurement band. In the last decade, the injection of squeezed states of light (or \textit{squeezing}) has been established as a promising technique to reduce quantum noise \cite{Eber, Stef, Vahl1, Chua, Vahl2, Khal, GEOSqz}, providing an opportunity to improve the detector sensitivity even further \cite{ET,US3G}. Due to the sub-attometer sensitivity Advanced LIGO aims to achieve, the interferometer needs to be carefully isolated from the outside world. To establish squeezing as a technology compatible with advanced gravitational wave detectors, it is critical to understand and quantify any potential noise coupling mechanism that could arise from squeezing injection.

One of the most pernicious enemies of gravitational-wave detectors operating at the quantum limit is scattered light \cite{Can, V1, V2, Ottaway,Frit}, i.e. light that scatters from a moving surface and reaches the interferometer readout photodetector. Depending on the scattered optical power and the scattering-object motion, scattered light can degrade the interferometer sensitivity, typically in the audio frequency region between 50 Hz and 300 Hz that is especially important for several astrophysical sources \cite{CutTho}. Backscattering noise is generally difficult to model as it depends on several variables which are not known {\it a priori}, such as the seismic motion transfer function to various optics and components of the interferometers. Many measurements of backscattered light impact have been made, focussing on the arm cavity beam tubes \cite{Tak,Ean1}, light baffles \cite{Ean2}, and in-air optical benches used for interferometer control \cite{TAcc}.

Squeezed state enhancement is achieved by injecting squeezed light into the output port of the Michelson \cite{Caves}, a separate location within an interferometer to the above-listed areas, and in the opposite propagation direction to outcoming interferometer optical beams. In the presence of squeezed state injection, the squeezing source itself (the Optical Parametric Oscillator (OPO)), becomes a scattering surface, causing scattered light to co-propagate with the squeezed vacuum state back towards the gravitational wave photodetector \cite{McK2,Vahl5,Frit}. Further, any scattered light circulating within the OPO is power-amplified by the optical parametric process that generates the squeezed state, of which the level of amplification is an unknown time-varying quantity. This makes an {\it a priori} estimate for the amount of backscattered light power and noise reaching the interferometer readout photodetector even more difficult.

In this paper, we report on the direct measurement of the impact of backscattered light from a squeezed-light source deployed on the 4 km LIGO H1 detector located in Hanford, WA, during the LIGO Squeezed Light Injection Experiment \cite{H1}. We also provide an analytical expression for the bidirectional scattering distribution function (BSDF) of an OPO. The techniques adopted to perform these measurements and the results obtained can be used to inform the design of a squeezing source for advanced detectors. We extrapolate our results to second generation advanced gravitational-wave detectors and their stricter requirements. More generally, these techniques can be applied to precision measurement experiments to assess the impact of backscattered light.

\section{Experiment overview and methods}
\begin{figure}[t]
\begin{center}
\includegraphics[width=10cm]{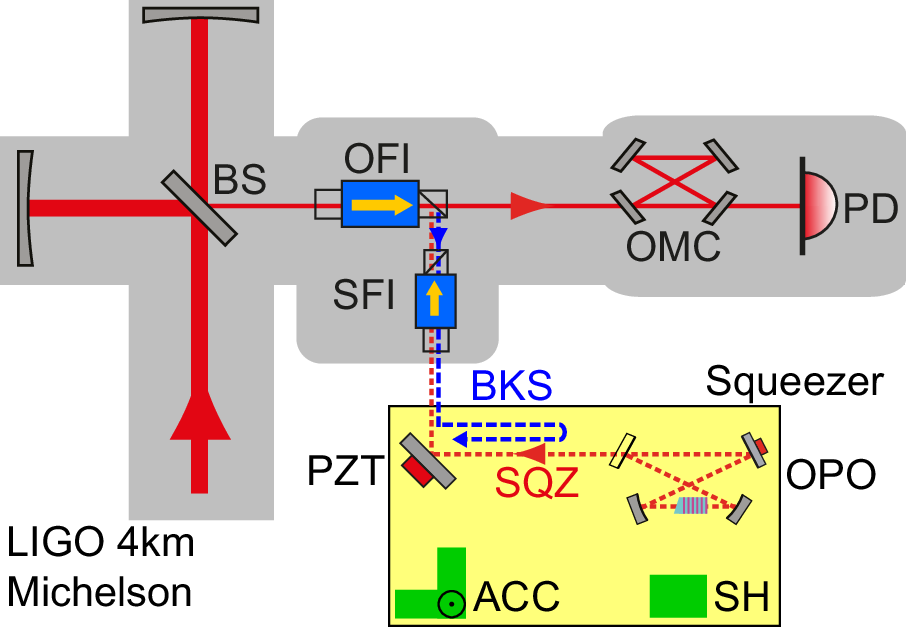}
\end{center}
\caption{Experiment overview (not to scale). Details of the squeezed-light source can be found in \cite{H1}. SQZ - squeezed light; OPO - squeezing generator cavity; BS - interferometer beamsplitter; OFI - output Faraday isolator; SFI - squeezing-injection Faraday isolator; OMC - output modecleaning cavity; PD - readout photodetector; Backscatter measurement hardware: PZT - injection path piezo-electric transducer; SH - shaker unit; ACC - accelerometers.}
\label{Schem}
\end{figure}

Figure \ref{Schem} shows a simplified schematic of the interferometer output chain and the squeezed-light source, or \textit{squeezer}. This includes the interferometer readout beam, or \textit{carrier}, output mode-cleaning cavity (OMC), readout photodetector (PD) and the output Faraday isolator (OFI) that isolates the Michelson from retro-reflections of the output chain optics.

The squeezer was situated on an optical table with tripod legs outside the vacuum envelope of the interferometer. Squeezing (SQZ, at the carrier wavelength $\lambda = 1.064\;{\rm \mu m}$), was injected through the OFI in reverse, to couple into the interferometer at the beamsplitter (BS). The measurement efficiency of squeezing at the interferometer output was $(38\pm2)\%$ during the backscattered light tests$^{\footnotemark}$\footnotetext{ A subsequently-improved squeezing measurement efficiency of $(44\pm 2)\%$ was reported in \cite{H1}.}. Our OPO consists of nonlinear periodically-poled potassium titanyl phosphate (PPKTP) crystal situated within a four-mirror traveling-wave optical cavity in bow-tie configuration. The intra-cavity beam waist located within the crystal was $W_{0}=34\;\mu{\rm m}$, and the input coupling mirror power reflectivity was $R_{\rm in} = (86.8\pm0.2)\%$. The OPO was specifically designed to isolate against backscattered light \cite{Chua}. For clarity, the other components of the squeezer, such as the pump field optics and hardware used to control the squeezing ellipse phase are not shown - further details of the squeezed-light source can be found in \cite{H1}.
\begin{figure}[t]
\center
\includegraphics[width=\textwidth]{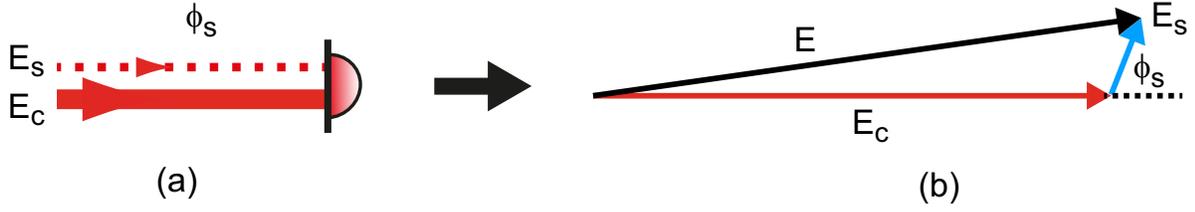}
\caption[Simple illustration and phasor diagram of beams incident on the readout photodetector]{(a) Simple illustration of beams incident on the readout photodetector. (b) Phasor diagram showing the relationships of the various fields.}
\label{scatterphasor}
\end{figure}
Light from the interferometer output port can be scattered by the OFI towards the squeezer. If a second scattering event from the squeezer table scatters this light back into the interferometer, it will form a backscattered light path (BKS), co-propagating with the injected squeezed light to the readout photodetector. An additional Faraday isolator (SFI) was installed in the squeezing injection path to reduce the scattered optical power. Placing extra Faraday isolators in the squeezing injection path could further increase the immunity to scattered light (thus reduce backscattered light), but would introduce additional optical losses that will limit the achievable squeezing improvement.

Here we introduce an expression for the noise due to incident backscattered light on the interferometer readout photodetector \cite{SChuaThesis}. We begin by denoting the carrier optical power as $P_{\rm c} = |E_{\rm c}|^{2}$, the backscatter optical power as $P_{\rm s} = |E_{\rm s}|^{2}\;(\ll P_{\rm c})$, and their relative phase as $\phi_{\rm s}$. The incident power on the interferometer readout photodetector is illustrated in figure \ref{scatterphasor}(a), along with the equivalent phasor diagram of the optical fields in \ref{scatterphasor}(b). The carrier power can be expressed as steady-state and fluctuating components $P_{\rm c} = \bar{P_{\rm c}}+\delta P_{\rm c}$, where $\delta P_{\rm c}\ll \bar{P}_{\rm c}$.

The total power detected by the readout photodetector, $P$, is given by
\begin{eqnarray}
P=E^{2}&=&E_{\rm c}^{2}+E_{\rm s}^{2}-2E_{\rm c}E_{\rm s}\cos(\pi-\phi_{\rm s})\nonumber\\
&=&P_{\rm c}+P_{\rm s}+2\sqrt{P_{\rm c}P_{\rm s}}\cos(\phi_{\rm s})\nonumber\\
&\approx&\bar{P}_{\rm c}+2\sqrt{\bar{P}_{\rm c}P_{\rm s}}\cos(\phi_{\rm s})\label{Peqn1}\\
&=&\bar{P}_{\rm c}+{\rm dS} \nonumber
\end{eqnarray}
where
\begin{equation}
{\rm dS} = 2\sqrt{\bar{P}_{\rm c}P_{\rm s}}\cos(\phi_{\rm s})\label{Peqn2}
\end{equation}
is the backscatter noise contribution in the interferometer readout. The relative phase $\phi_{\rm s}$ is assumed to accrue entirely from the total beam-path displacement due to the scattering object $Z$, via the relation
\begin{equation}
\phi_{\rm s} = 2kZ = \frac{4\pi}{\lambda}Z
\end{equation}
with the wavenumber $k=2\pi/\lambda$. The total beam-path displacement can be written as two terms that describe the contributions from large beam-path displacements, $Z_{\rm s}$, and small beam-path displacements, $\delta z_{\rm s} (\ll Z_{\rm s})$, namely $Z = Z_{\rm s}+\delta z_{\rm s}$.

Equation (\ref{Peqn2}) can now be expressed in terms of {\it relative intensity noise} (RIN), calculated from the fluctuating component of the total power normalized by the average total power, namely
\begin{eqnarray}
{\rm RIN_{s}} = \frac{{\rm dS}}{\bar{P}_{c}} &=& 2\sqrt{\frac{P_{\rm s}}{\bar{P}_{c}}}\cos(\phi_{s})\nonumber\\
&=&2\sqrt{\frac{P_{\rm s}}{\bar{P}_{c}}}\cos(2k(Z_{\rm s}+\delta z_{\rm s}))\nonumber\\
&\approx&2\sqrt{\frac{P_{\rm s}}{\bar{P}_{c}}}\left[\cos(2kZ_{\rm s})\cos(2k\delta z_{\rm s})+\sin(2kZ_{\rm s})\sin(2k\delta z_{\rm s})\right]\label{Peqn3}
\end{eqnarray}

We further simplify equation (\ref{Peqn3}) by approximating over many cycles $\sin(2kZ_{\rm s})\approx1/\sqrt{2}$ and $\cos(2k\delta z_{\rm s})\approx1$, resulting in:
\numparts
\begin{eqnarray}
{\rm RIN_{s}}&\approx2\sqrt{\frac{P_{{\rm s}}}{\bar{P}_{c}}}\cos{\left(\frac{4\pi Z_{\rm s}}{\lambda}\right)}\label{RINsa}\\
&\quad+\sqrt{\frac{2P_{\rm s}}{\bar{P}_{c}}}\left(\frac{4\pi\delta z_{\rm s}}{\lambda}\right)\label{RINsb}
\end{eqnarray}
\endnumparts
This shows the backscatter noise contribution in the interferometer readout is separable into large displacement (\ref{RINsa}) and small displacement terms (\ref{RINsb}), and that it is dependent on the amount of (DC) optical power of the backscattered beam $P_{\rm s}$, and the motion of the scattering object $Z = Z_{\rm s} + \delta z_{\rm s}$.

During normal squeezed-interferometer operating conditions, the sensitivity of the interferometer was broadly enhanced with injected squeezed light \cite{H1} - backscattered light reflected from the squeezer did not degrade the sensitivity at any frequency. To characterise the backscattered light impact we therefore intentionally applied displacement motion to induce a backscattered light response in the readout. Figure 1 also shows the hardware used for backscattered light measurements. These were a piezo-electric transducer (PZT) in the squeezing injection path, a piezo-driven shaker unit (SH), and three orthogonally-mounted accelerometers (ACC).

All backscattered light measurements reported in this paper were undertaken with squeezed light being injected into the interferometer. Applying displacement motion adds phase to the squeezing injection path, which would rotate the squeezing ellipse orientation, potentially causing antisqueezing to enter the interferometer readout. However, the squeezing ellipse phase angle control loop, with a control bandwidth of 10s of kHz and much higher range than the applied displacement motion magnitudes and frequencies, maintained the squeezing ellipse phase angle so that squeezing was matched to the interferometer readout during the tests.

Two tests of applied displacement motion were undertaken, each in different motion magnitude regimes - large motion $Z_{\rm s}$ and small motion $\delta z_{\rm s}$. This allows us to use equations (\ref{RINsa}) and (\ref{RINsb}) in determining different backscattering characteristics of the system, namely, the backscattered light power reaching the photodetector, the backscatter reflectivity parameter of the OPO, and the level of background backscatter noise.

\section{Large displacement motion}\label{Largem}
Nonlinear noise up-conversion \cite{V1} of backscattering noise can result from displacement motion of the order of one wavelength ($Z_{\rm s}\sim\lambda$). We intentionally applied large sinusoidal displacement motion to the backscatter optical path length using the injection-path PZT. The up-conversion appears as a ``shelf'' signal in the readout spectrum \cite{Ottaway}. The shelf structure characteristics reflect the applied displacement motion and the backscattered light condition, and is encapsulated by equation (\ref{RINsa}). Of specific interest, the shelf plateau height is proportional to the backscatter-to-carrier power ratio ($P_{{\rm s}}/\bar{P}_{c}$). With known applied modulation parameters and the carrier power $\bar{P}_{c}$, we can infer the backscattered light power reaching the photodetector $P_{{\rm s}}$.

\begin{figure}[t]
\begin{center}
\includegraphics[width=10cm]{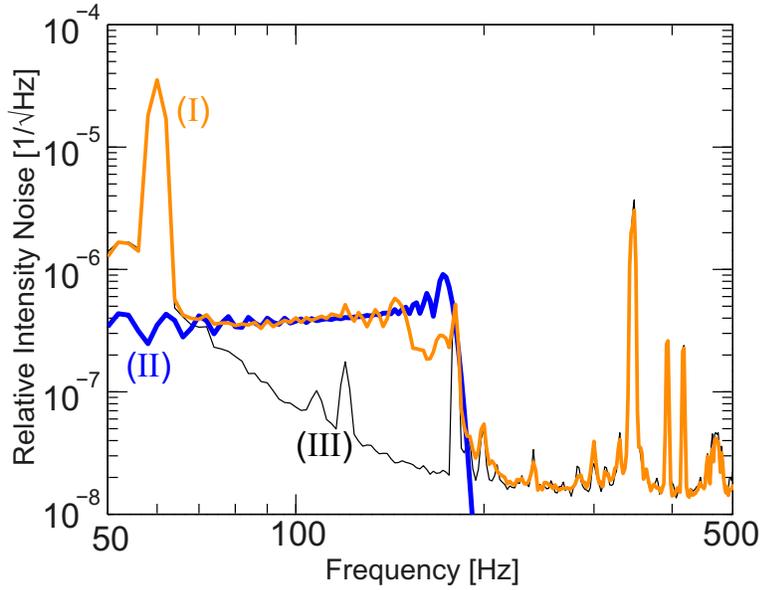}
\end{center}
\caption{(I) Interferometer readout RIN with induced shelf-structure from applied known backscatter pathlength modulation; (II) equation (\ref{RINsa}) model, using known parameters and fitting for $P_{{\rm s}}/\bar{P}_{c}$; (III) Interferometer total noise with no modulation.}
\label{FWplot}
\end{figure}

With squeezed light being injected and applied PZT motion (drive frequency of 1 Hz and modulation depth of 173 rad), the induced shelf structure observed is shown in figure \ref{FWplot}. Using equation (\ref{RINsa}) and the known modulation parameters of the sinusoidal drive signal, fitting to the measured spectrum determined the backscatter-to-carrier power ratio to be $P_{{\rm s}}/\bar{P}_{c} = (1.7\pm0.2)\times10^{-11}$. The backscatter power of the squeezer reaching the interferometer readout photodetector is therefore determined to be $P_{\rm s} = (260\pm40)\;{\rm fW}$, calculated using the measured carrier power $\bar{P}_{c} = 16.1\;{\rm mW}$.

\subsection{Backscatter Reflectivity of the OPO}\label{BackReflOPO}
The squeezed-light-generating OPO is the dominant backscattering optic within the squeezed light injection path. The back reflectivity of the OPO to scattered light, $R_{\rm OPO}$, is a valuable parameter in evaluating the backscattering immunity of the squeezer. We can use the determined backscatter power value $P_{\rm s}$ above to infer $R_{\rm OPO}$.

The backscatter reflectivity parameter is the ratio of backscatter power being reflected by the OPO to the spurious scattered power incident on the OPO:
\begin{itemize}
\item The backscatter power being reflected at the OPO is the measured backscatter power value $P_{\rm s}$ corrected for the measurement efficiency of light from the OPO reaching the readout photodector $\eta = (38\pm2)\%$.
\item The spurious scatter power incident on the OPO is the carrier frequency light power scattered towards the squeezer, measured to be $P_{\rm sp} = (0.7\pm0.1)\mu{\rm W}$, corrected for the proportion matching the spatial and polarization mode of the OPO, measured to be $\rho = (11\pm3)\%$.
\end{itemize}
Therefore the backscatter reflectivity parameter is given by:
\begin{equation}
R_{\rm OPO} = \frac{P_{\rm s}}{\eta \rho P_{\rm sp}}
\end{equation}
We calculate that $R_{\rm OPO} = -(50\pm1)\;{\rm dB}$, that is, the traveling-wave design of our OPO provides $(50\pm1)$ dB of intrinsic isolation to backscattered light. This is $\sim$20 dB greater isolation than typical off-the-shelf Faraday isolators.

Further, we can use the $R_{\rm OPO}$ value to calculate the backscattering distribution function of our OPO. We assume that the dominant source of scatter is the nonlinear crystal, and that the input coupler mirror of the OPO is the dominant source of optical transmission loss in the optical cavity. We also account for the parametric process within the OPO, that amplifies or de-amplifies the scattered light depending on the relative phase between the circulating scattered light and the OPO pump field. The backscatter reflectivity can now be written as \cite{SDwyerThesis}:
\begin{equation}
R_{\rm OPO} = {\rm BSDF}\times\frac{16\Omega_{1/e}}{(1-R_{\rm in})^{2}}\left(\frac{1-2x\cos{\theta_{\rm sc}}+x^{2}}{(1-x^{2})^{2}}\right)\label{ROPO2}
\end{equation}
where the bidirectional scattering distribution function (BSDF) is a measure that characterizes the backscattering impact for a transmissive optic, such as in the case of the nonlinear crystal of the OPO. We can infer the BSDF for the nonlinear crystal using the determined backscatter reflectivity value $R_{\rm OPO}$ \cite{Frit}.

In equation (\ref{ROPO2}), $\theta_{\rm sc}$ is the relative phase between the carrier light scattered into the OPO and the phase of the circulating OPO pump field. This is related to the squeezing ellipse angle but also depends on the distance between the OPO and the interferometer, which changes slowly on the scale of microns. $\theta_{\rm sc}$ is not known accurately and can change between measurements. We therefore make the assumption that $\theta_{\rm sc} = 0$, which will give the highest (and most conservative) BSDF result.

The other parameters in equation (\ref{ROPO2}) are the input coupler power reflectivity $R_{\rm in} = (86.8\pm0.2)\%$, the normalized parametric interaction strength of the OPO in this experiment $x = 0.6$, and the solid angle at the crystal $\Omega_{1/e} = \lambda^{2}/(\pi W_{0}^{2})$,  ($\lambda = 1.064\;\mu{\rm m}$, $W_{0} = 34\mu{\rm m}$) \cite{Frit, Raab, SDwyerThesis}. Using the determined $R_{\rm OPO}$ and equation (\ref{ROPO2}), this results in an inferred BSDF for the crystal of $(9\pm3)\times10^{-5}\;{\rm str}^{-1}$. This BSDF value is comparable in magnitude to measured BSDF values of off-the-shelf single optical components \cite{Zucker}. The inference of the BSDF value provides options to further improve the backscattered light immunity of the OPO. We discuss these possible pathways in the final section of the paper.

\section{Small displacement motion}
Backscatter signals from motions of magnitude much smaller than the optical wavelength ($\delta z_{{\rm s}} \ll \lambda$) couple linearly to the readout spectrum \cite{V2,Ottaway}. From this linearity and equation (\ref{RINsb}), we can derive a relative relationship between background (`bg') and driven (`dr') measurements of backscatter RIN and squeezer table motion. This allows us to infer the background backscatter noise level that impacts the interferometer readout, that is:
\begin{equation}
{\rm RIN}_{\rm s-bg}={\rm RIN}_{\rm s-dr}\times\frac{\delta z_{\rm s-bg}}{\delta z_{\rm s-dr}}\label{linear}
\end{equation}
Therefore, by applying small displacement motion, this allows us to infer the backscattered noise level at measurement frequencies of interest, particularly in the most sensitive band of the interferometer readout.

We induced small displacement motion on the squeezer by using a shaker unit, and measuring the motion with the accelerometers mounted to the optical table$^{\footnotemark}$\footnotetext{Ideally, a shaker unit should also be used in the large displacement motion measurement, to closer-reflect environmental vibration behaviour on the apparatus. However, the shaker unit available did not have enough drive-range to actuate in the large-motion regime.}. With squeezed light being injected into the interferometer, measurements of the backscattering noise induced into the interferometer spectrum (${\rm RIN}_{\rm s-dr}$), and measurements of background and driven table motion (${\delta z}_{\rm s-bg},\;{\delta z}_{\rm s-dr}$) were made.

\begin{figure}[t]
\begin{center}
\includegraphics[width=10cm]{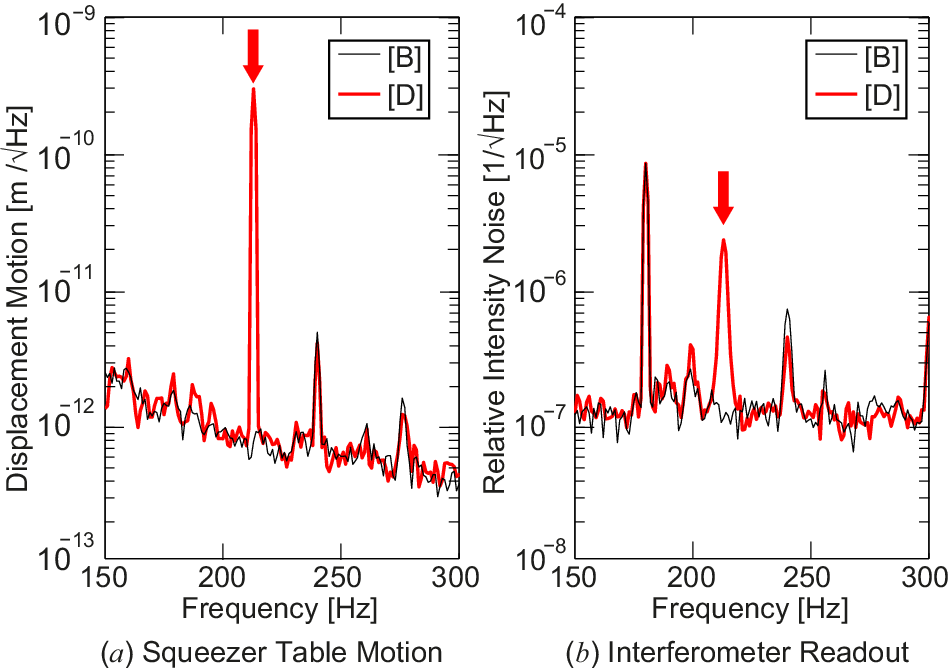}
\end{center}
\caption{Background motion and driven shaker unit measurements, corresponding to traces `[B]' and `[D]' respectively. Single-frequency excitation driven above background levels are indicated by the arrows. $(a)$ Squeezer table motion spectrum and $(b)$ RIN spectrum at the interferometer readout photodiode.}
\label{ShakerSingle}
\end{figure}

Firstly, the linear-coupling condition for small displacement motion was tested using the shaker and accelerometers in a separate measurement run. A single-frequency displacement motion at 270 Hz was applied to the squeezer table, and the shaker drive amplitude was varied by a factor of 10. After applying equation (\ref{linear}), the resulting inferred  background backscatter noise level clustered to within a factor of 2 of each other for all drive amplitudes used. This provided confidence with the linearity condition.

\begin{figure}[t]
\begin{center}
\includegraphics[width=10cm]{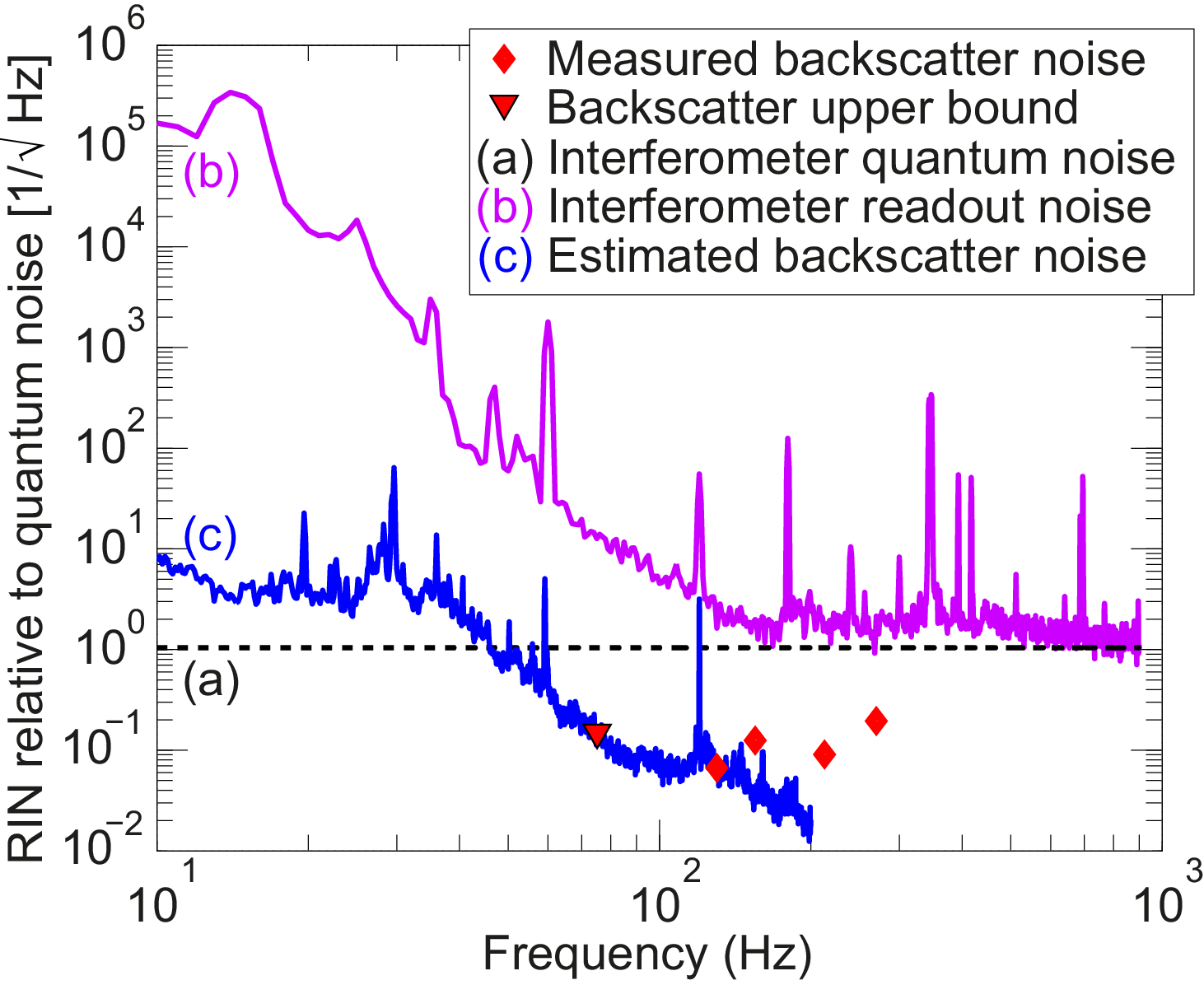}
\end{center}
\caption{Backscatter RIN and Advanced LIGO requirement levels. All data and traces are normalized to the H1 interferometer quantum noise level (c). The 75 Hz upper-limit point is denoted separately.}
\label{Shaker}
\end{figure}

Then single-frequency excitations were made for several different frequencies - figure \ref{ShakerSingle} shows one such single-frequency measurement. These frequencies were chosen so as to fall within regions of the interferometer readout spectrum with no nearby peaks. Using equation (\ref{linear}), we then inferred the level of background backscatter noise. The inferred backscatter noise level for several injections is shown by the data points in figure \ref{Shaker}. All data and traces are normalized to the H1 interferometer quantum noise level - trace (a). Below 300 Hz, the noise level due to background backscattered light is around a factor of 10 below the interferometer readout noise (trace (b) in figure \ref{Shaker}), and at least a factor of 7 below the quantum noise level. The 75 Hz point is denoted separately as it represents an upper limit, with induced scatter noise not measurable in the interferometer readout with the applied motion.

As further verification, we compare the single-frequency measurement result to an estimate of the background backscatter noise level, given the amount of backscattered light power reaching the interferometer readout photodetector $P_{\rm s}$ and the background table motion $\delta z_{\rm s-bg}$. Using equation (\ref{RINsb}), the backscatter noise ${\rm RIN_{s-bg}}$ relative to interferometer quantum noise ${\rm RIN_{\rm qn}} = \sqrt{2hc\lambda/\eta\bar{P}_{\rm c}}$, can be written as \cite{Frit}:
\begin{equation}
\frac{\rm RIN_{\rm s-bg}}{\rm RIN_{\rm qn}} = (4 \pi \delta z_{\rm s-bg}) \sqrt{\frac{\eta P_{\rm s}}{\lambda h c}}
\end{equation}
where $\eta$ is the photodiode efficiency, $h$ is the Planck constant, and $c$ is the speed of light. Given the background motion of the squeezer table measured by the accelerometers, photodiode efficiency of $(96\pm2)\%$, and $P_{\rm s} = (260\pm40)\;{\rm fW}$ determined from the large-motion nonlinear upconversion measurement in section \ref{Largem}, the expected background backscatter noise contribution is shown by trace (c). The single-frequency backscatter measurements are in line with the backscatter noise contribution expected from the amount of backscattered light $P_{\rm s}$ that reaches the readout photodetector. Above 150 Hz, the deviation is attributed to  mechanical resonances in the optical set-up that might amplify the applied table motion.

\section{Implications for future interferometers}
Future generations of interferometers are expected to be limited by quantum noise across their entire measurement frequency band, and squeezed state injection is expected to be included in baseline technology for such instruments \cite{ET,US3G}. In this final section, we extrapolate the presented backscatter measurement results for the second generation instrument of LIGO, known as Advanced LIGO.

\begin{figure}[t]
\begin{center}
\includegraphics[width=10cm]{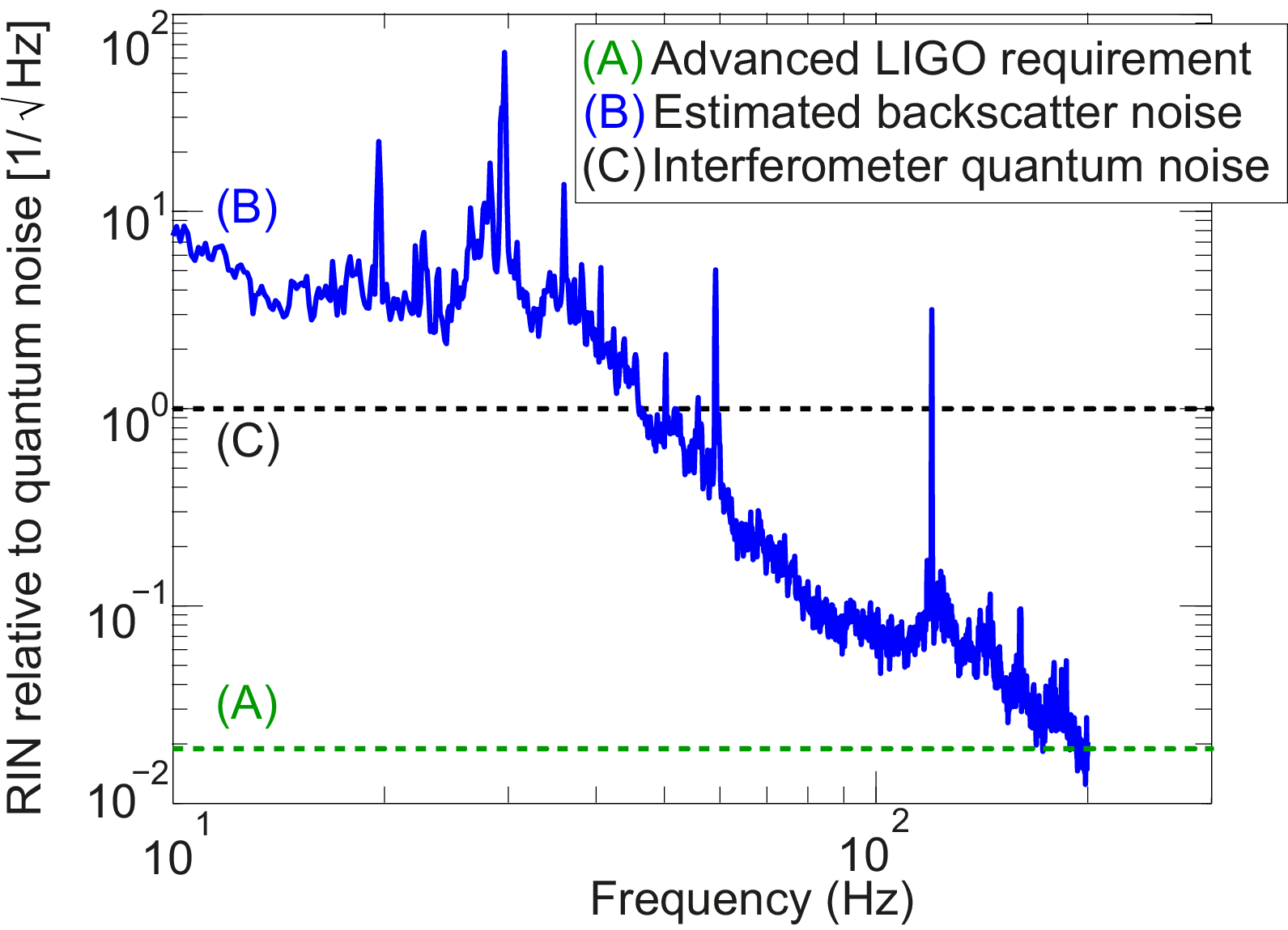}
\end{center}
\caption{The estimated backscatter RIN level and Advanced LIGO requirement levels. The estimated backscatter RIN level trace (B), is the same as trace (b) of figure \ref{Shaker}. All data and traces are normalized to the H1 interferometer quantum noise level (trace (C)).}
\label{AdvLIGOReq}
\end{figure}

We first consider the requirements for backscatter noise compared to the Advanced LIGO design sensitivity (reported in \cite{Harry}). Quantum noise will be a limiting noise source across the entire measurement frequency band, and we require scattering noise to be at least a factor of 10 below quantum noise. Due to the expected operating state of the interferometer and increased input laser power, a greater amount of carrier light power is expected to leave the interferometer beamsplitter toward the OFI, an increase of up to a factor of 7 \cite{Harry}. Consequently, more spurious light could potentially enter the squeezed beam path towards the OPO. Moreover, injected squeezing will further reduce quantum noise as much as 6 dB \cite{H1}, therefore we impose another factor of 2 in the requirements. All these requirements are encapsulated by trace (A) of figure \ref{AdvLIGOReq}.

Comparing the Advanced LIGO requirement (trace (A)) to the estimated backscattered light level (trace (B), the same as trace (b) of figure \ref{Shaker}), the conclusion is that the current set up for backscattering mitigation will not be adequate for the Advanced LIGO requirements. A factor of 10 additional isolation from backscatter noise is needed above 60 Hz, and an additional isolation factor of about 400 is needed below 40 Hz. There are various options, available to reduce backscatter noise to the requirement level.

Improvement in the seismic isolation of squeezer is a viable option to reduce the impact of backscattered light. For instance, moving the OPO onto an Advanced LIGO seismic isolation platform within the interferometer vacuum envelope will provide an additional isolation factor of $\sim$3000 at 10 Hz \cite{Harry} to ground motion. This is sufficient to reduce the background backscatter noise below the full Advanced LIGO requirement level. As the Advanced LIGO seismic isolation platforms are now being constructed as part of the instrument upgrade to Advanced LIGO, the technology is now becoming available, thus an increase in seismic isolation for the OPO is a very feasible prospect.

Another pathway for increasing immunity to backscattered light is to reduce the level of spurious light that reaches the squeezed-light source, reducing the amount of power that can be backscattered. This reduction can be made by increasing the isolation of the single squeezing path Faraday isolator. This is a continuing research area for improving Faraday materials and construction.

From section \ref{BackReflOPO}, a reduction of the impact of backscattered light can also be achieved by reducing the backscatter reflectivity parameter $R_{\rm OPO}$. Examining equation (\ref{ROPO2}), the backscatter reflectivity parameter could be reduced by (i) reducing the BSDF of the OPO, (ii) increasing the power transmission of the input coupler or (iii) by increasing the size of the cavity waist at the crystal, or a combination of the three options. The optics used to form the OPO had the best available polishing and surface coatings at the time, thus the pursuit of reducing the BSDF is a long-term research area for optical polishing and coating technologies.

The effect of either option (ii) and (iii) will be to change the input operating parameters of the OPO, in particular, increase the amount of pump power needed to drive the OPO nonlinear process to reach a given squeezing generation level. There is a trade off between further increasing backscatter immunity and keeping the operating parameters of the squeezing OPO at functional levels. However, these two options (or a combination of both) can be currently achieved with careful choice of OPO mirror characteristics. For example, a decrease of the input coupler reflectivity to $R_{\rm in}=80\%$ will decrease the $R_{\rm OPO}$ value by a factor of 2.

Finally, we explicitly note that combinations of the above options will cascade the individual improvements together. These combinations would provide the greater backscattered light noise suppression margins that will become necessary for future interferometers with greater sensitivities and even more stringent requirements.

In this paper we have presented results quantifying the backscattered-light impact in squeezing-enhanced gravitational-wave detection. We have demonstrated methods to quantify both the level of backscattered light power, the backscatter reflectivity parameter of the OPO, and the level of backscatter noise it introduces. Our results show that backscattered light is a surmountable technical challenge to the use of squeezed states in future gravitational-wave detectors.

\ack{The authors thank the staff at the LIGO Hanford Observatory for invaluable help to facilitate the experiment. In particular, we are grateful to Carol Wilkinson who had the difficult task of keeping the Advanced LIGO project on schedule, and to David Barker and Richard McCarthy who maintained the H1 computer and electronics systems. LIGO was constructed by the California Institute of Technology and Massachusetts Institute of Technology with funding from the National Science Foundation and operates under cooperative agreement PHY-0757058. This paper has the LIGO document number LIGO-P1200155-v9.}

\section*{References}
\bibliographystyle{unsrt}
\bibliography{ChuaBibliography}

\end{document}